\documentclass[sigconf]{acmart}
\usepackage{hyperref}
\usepackage{url}
\usepackage{multirow}
\usepackage{booktabs}
\usepackage{subcaption}
\usepackage{float}   
\usepackage{adjustbox} 
\usepackage{wrapfig} 
\usepackage{caption} 
\usepackage{siunitx}
\AtBeginDocument{%
  }

\setcopyright{acmlicensed}

\copyrightyear{2026}
\acmYear{2026}
\acmConference[SIGSPATIAL '26]{The 34th ACM International Conference on Advances in Geographic Information Systems}{November 03--06, 2026}{Riverside, CA, USA}
\acmBooktitle{The 34th ACM International Conference on Advances in Geographic Information Systems (SIGSPATIAL '26), November 03--06, 2026, Riverside, CA, USA}
\acmDOI{10.1145/3841645.3843304}
\acmISBN{979-8-4007-2950-8/2026/11}

\begin{document}

\title{A-UTE: Advection Informed, Uncertainty Aware Temperature Emulator} 

\author{Hira Saleem}
\email{h.saleem@unsw.edu.au}
\orcid{0009-0000-9046-0446}
\affiliation{%
  \institution{Department of Computer Science and Engineering  University of New South Wales}
  \city{Sydney}
  \state{NSW}
  \country{Australia}
}

\author{Cormac Purcell}
\email{cormac.purcell@unsw.edu.au}
\orcid{0000-0002-7491-7386}
\affiliation{%
  \institution{Department of Computer Science and Engineering University of New South Wales}
  \city{Sydney}
  \state{NSW}
  \country{Australia}
}

\author{Steven Sherwood}
\email{s.sherwood@unsw.edu.au}
\orcid{0000-0001-7420-8216}
\affiliation{%
  \institution{Climate Change Research Centre \\
  University of New South Wales}
  \city{Sydney}
  \state{NSW}
  \country{Australia}
}

\author{Flora Salim}
\email{flora.salim@unsw.edu.au}
\orcid{0000-0002-1237-1664}
\affiliation{%
  \institution{Department of Computer Science and Engineering University of New South Wales}
  \city{Sydney}
  \state{NSW}
  \country{Australia}
}

\renewcommand{\shortauthors}{Saleem et al.}

\begin{abstract}
Spatially resolved monthly temperature projections are an essential data product for climate impact assessment and scenario analysis, however producing them with physics-based Earth system models (ESMs) is computationally expensive, limiting the ensembles and scenarios that can be explored. We present A-UTE, an advection-informed, uncertainty-aware temperature emulator: an application of spatial machine learning that forward-integrates gridded monthly-mean surface air temperature. Given the initial condition of temperature field and prescribed monthly forcing maps ($CO_2, CH_4, SO_2$, and black carbon), A-UTE auto-regressively predicts 10-year horizons. The architecture pairs a 2D forced-advection ODE with a latent neural SDE residual. Trained via a latitude-weighted Gaussian negative log-likelihood, A-UTE outputs predictive means and calibrated grid-level uncertainties. We train and evaluate on simulation output from 20 physics-based climate models under SSP forcing scenarios: 19 CMIP6 ESMs from ClimateSet ($\sim$250\,km grid) and FV3GFS ($1^\circ$ grid), training on SSP1-2.6 and testing on the unseen SSP2-4.5 scenario. A-UTE achieves lower latitude-weighted RMSE and CRPS than U-Net, ConvLSTM, SFNO, and a graph-based advection-diffusion baseline, with bounded error across 120-step roll-outs, and transfers zero-shot to ESMs unseen during training. A-UTE offers a practical route to large, scenario-conditioned probabilistic ensembles of gridded temperature for downstream spatial analysis while running at a small fraction of the cost of an ESM integration.
\end{abstract}





\maketitle

\section{Introduction}
Spatially explicit projections of future temperature are increasingly used to support climate-aware decision making across sectors. Several applications ranging from infrastructure planning and energy-system management to regional heat-risk analysis depend on gridded estimates of how temperature evolves under alternative emission pathways \cite{meehl2021initialized, troccoli2010seasonal}. These projections are typically generated by physics-based Earth system models (ESMs), which evolve the coupled atmosphere-ocean system under prescribed forcings to produce physically consistent, scenario-dependent responses \cite{eyring2016overview}. The value of these projections for downstream decision-making rests not only on their physical realism, but also on the ability to characterise uncertainty through ensembles spanning different initial conditions, forcing pathways, and model structures. Generating such ensembles remains computationally demanding. CMIP6-class ESMs require thousands of CPU-hours per simulated year \cite{balaji2017cpmip}, constraining the number of scenarios, members, and configurations that can be explored. Fast data-driven emulators that reproduce ESM behaviour under prescribed forcings offer a practical route to expanding ensemble size and diversity at a fraction of the computational cost \cite{tebaldi2025emulators}.

This work focuses on spatially resolved monthly-mean near-surface air temperature (TAS). Temperature is the most widely used climate variable in impact assessment and adaptation planning, serving as a direct input to heat risk analysis, degree-day calculations, agricultural assessments, energy-demand modelling, and ecosystem studies. It is also among the most consistently archived outputs across ESM ensembles, making it a natural target for large-scale emulation efforts \cite{beusch2020emulating, nath2022mesmer}. Rather than emulating the full Earth system state, we study the evolution of a single, high-value geospatial field. Given a monthly global temperature map and prescribed radiative forcing maps, we predict the temperature field at the next monthly timestep. This framing supports rigorous evaluation of spatial accuracy, uncertainty quantification, cross-scenario generalisation, and transfer across climate models while addressing a climate data product that is widely consumed by downstream applications. We approach the problem as a spatio-temporal machine-learning task, learning the evolution operator of a gridded geophysical field with explicit attention to spatial structure, transport-informed dynamics, and per-grid-cell predictive uncertainty.

\begin{figure}
    \centering
      \includegraphics[scale=0.3]{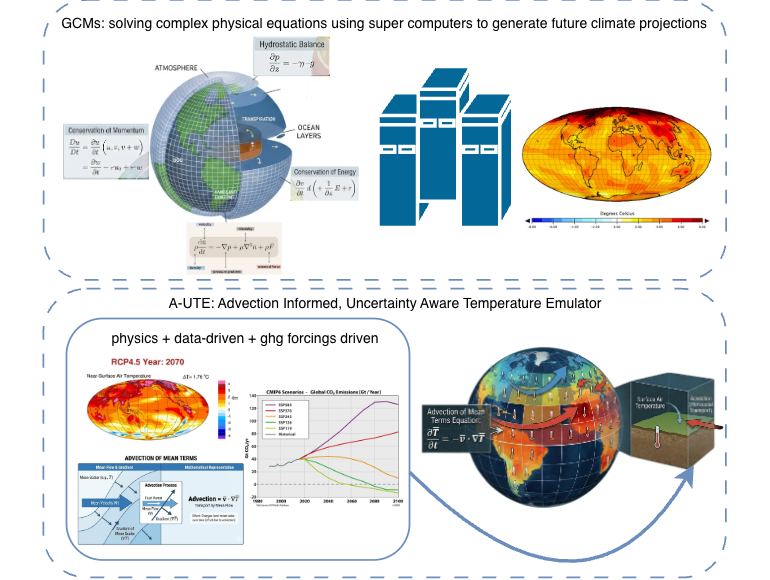}
    \caption{GCM vs. A-UTE (emulation). General Circulation Models (GCMs) numerically solve the coupled atmosphere  physics to simulate climate evolution under prescribed forcings, producing physically grounded trajectories at high computational cost. A-UTE is a hybrid data-driven emulator, it uses a forced advection-based dynamical core to provide a coarse-grained transport prior, and a latent Neural SDE residual trained with a likelihood objective to produce calibrated probabilistic roll-outs at a fraction of the runtime, enabling efficient long-horizon scenario analysis and cross-model evaluation.}
    \label{gcmvsaute}
\end{figure}

Current approaches to climate emulation broadly fall into two distinct families. Statistical emulators, notably the MESMER family, map global-mean trajectories to spatially resolved temperature realizations through engineered parametric forms. These models are mature, highly interpretable, and widely integrated into climate impact pipelines \cite{beusch2020emulating, nath2022mesmer}. In contrast, deep-learning dynamical emulators integrate the atmospheric state forward in time, achieving stable, high-skill roll-outs \cite{watt2023ace, guan2024lucie}. However, these are typically constrained to training on a single Earth System Model (ESM) at sub-daily resolutions. A critical gap therefore remains. An emulator that is spatially resolved, scenario-conditioned, and probabilistic, yet trained across heterogeneous ESMs and grid resolutions—would fundamentally extend the capabilities of both families, particularly by operating directly on the monthly-aggregated temperature fields standard in impact applications.

Embedding physically motivated spatial structure into a data-driven operator can improve stability and generalisation. Temperature evolution exhibits coherent spatial propagation patterns across neighbouring regions, arising from the combined influence of atmospheric circulation, large-scale climate variability, land-ocean contrasts, and externally forced responses. Rather than modelling these mechanisms explicitly, A-UTE introduces a transport-informed spatial prior that encourages the learned operator to represent how temperature patterns evolve across space. Specifically, A-UTE integrates a two-dimensional forced-advection ODE driven by a pseudo-velocity field derived from the temperature data itself. The pseudo-velocity is not intended to represent the physical near-surface wind field, instead, it provides an effective description of spatial transport patterns present in the training data. This transport-shaped prior introduces a structured inductive bias for spatial evolution, while a neural stochastic residual component allows the model to capture additional variability and uncertainty present in the monthly climate fields. As shown in Section~\ref{sec:ablation}, incorporating this prior consistently improves predictive skill across ESMs and scenarios.

A-UTE (\emph{Advection-informed, Uncertainty-aware Temperature Emulator}) proposes this design as a two-stage operator. Given an initial monthly TAS field and a sequence of prescribed monthly forcing maps (CO\textsubscript{2}, CH\textsubscript{4}, SO\textsubscript{2}, and black carbon), it auto-regressively predicts the next monthly field. A forced advection ODE supplies the transport-shaped spatial prior, while a latent neural SDE refines the trajectory and models stochastic variability. Training under a latitude-weighted Gaussian negative log-likelihood yields a predictive mean together with a calibrated, heteroscedastic uncertainty estimate at every grid cell. We train and evaluate on simulation output from 20 physics-based climate models under SSP forcing scenarios-19 CMIP6 ESMs from ClimateSet \cite{kaltenborn2023climateset} at $\sim$250,km resolution and FV3GFS \cite{zhou2019toward} at $1^\circ$ resolution-training on SSP1-2.6 and testing on the held-out SSP2-4.5 scenario. Our contributions are:

\begin{itemize}
\item \textbf{A transport-informed probabilistic emulator for gridded monthly temperature.}
We couple a forced-advection ODE, acting as a transport-shaped spatial prior over monthly temperature fields, with a latent neural SDE that refines the predicted trajectory, yielding an operator that is both physically motivated and probabilistic.

\item \textbf{Calibrated, spatially resolved uncertainty.}
We train under a latitude-weighted Gaussian NLL with multiple Brownian realisations, producing per-grid-cell predictive variance and enabling probabilistic evaluation (CRPS) alongside point accuracy.

\item \textbf{Stable, multi-model, scenario-conditioned evaluation.}
Across 20 ESMs at two grid resolutions, A-UTE achieves lower
latitude-weighted RMSE and CRPS than U-Net, ConvLSTM, SFNO, and a graph-based advection-diffusion baseline, with error remaining bounded over 120-step (10-year) roll-outs, while transferring zero-shot to ESMs unseen during training at a small fraction of the cost of ESM integration.
\end{itemize}

\section{Related Work}
\label{sec:related}
Physics-based climate models solve discretized governing equations on global grids and remain computationally expensive for long integrations, motivating data-driven surrogates that accelerate experimentation. Recent climate emulators largely fall into two settings: purely data-driven or hybrid physics based emulators. Large-scale datasets such as ClimateSet \cite{kaltenborn2023climateset}, built from inputs/outputs of many CMIP6/Input4MIPs models, further enable multi-model benchmarking and the development of data-driven emulators trained across heterogeneous sources.

\paragraph{Statistical emulators of spatial temperature.}
The most established route to spatially resolved temperature emulation is the MESMER family of statistical emulators. MESMER generates grid-cell-level annual temperature realisations conditioned on global-mean trajectories \cite{beusch2020emulating}. MESMER-M extends this to monthly resolution \cite{nath2022mesmer}, MESMER-X to spatially resolved annual extremes \cite{quilcaille2022showcasing}, and MESMER-M-TP to joint monthly precipitation and temperature \cite{schongart2024introducing}. These emulators combine parametric forced-response models with auto-regressive variability components and spatially correlated stochastic generators, and pattern-scaling approaches similarly map global-mean warming to local responses through scenario-independent spatial patterns \cite{mansfield2020predicting}. Their low parameter count and interpretability have made them a standard component of impact pipelines, and they operate on hand-specified parametric forms conditioned on a global-mean trajectory \cite{tebaldi2025emulators}.

\paragraph{Deep-learning based emulators.}
A parallel line of work emulates atmospheric dynamics directly with neural networks that predict the next atmospheric state from the current one, building on data-driven medium-range weather models \cite{lam2023learning, kochkov2024neural, nguyen2023climax}. The Ai2 Climate Emulator and its successor ACE2 integrate global state at sub-daily timesteps and reproduce subseasonal-to-decadal variability and forced responses \cite{watt2023ace, watt2025ace2}, and have been transferred to other atmospheric models and to slab-ocean configurations sensitive to CO\textsubscript{2} \cite{duncan2024application, clark2025ace2}. LUCIE achieves stable roll-outs for ensembles of thousands of members \cite{guan2024lucie, guan2025lucie}. Probabilistic variants such as Spherical DYffusion \cite{cachay2024probabilistic} and ArchesClimate \cite{clyne2025archesclimate} produce ensembles through conditional generative formulations. These models establish that stable long-horizon integration is achievable and are typically trained on a single ESM and operate on instantaneous sub-daily state \cite{chattopadhyay2023long}.

DiffESM uses three-dimensional conditional diffusion to emulate monthly temperature and precipitation \cite{bassetti2024diffesm}, and score-based generative models have been applied to impact-relevant ESM outputs conditioned on global-mean temperature \cite{bouabid2026score}. These approaches generate the aggregated field through diffusion-based inference and are typically demonstrated at single-ESM scope.

\paragraph{Physics-structured emulation and operator learning.}
A complementary direction encodes partial physical structure as an inductive bias on the learned operator. The neural advection-diffusion equation (NADE) couples a learned advection-diffusion form on a graph with a neural correction on regional climate data \cite{choi2023climate, hwang2021climate}, and differentiable atmospheric cores embed conservation structure inside hybrid general circulation models \cite{kochkov2024neural}. This strategy draws on a broader toolkit for learning spatiotemporal dynamics. Neural operators that learn solution maps for parametric PDEs \cite{li2020fourier, lu2021learning, bonev2023spherical}, physics-informed networks for advection-dominated transport \cite{he2021physics, vadyala2022physics}, message-passing PDE solvers \cite{brandstetter2022message}, and continuous-time formulations based on neural ordinary and stochastic differential equations \cite{chen2018neural, kidger2021neural, shen2025neural}. 

While existing emulators demonstrate stable long-horizon roll-outs and probabilistic ensemble generation, there remains room for methods that incorporate an explicit physics-consistent prior for the coarse-grained dynamics of the emulated variable (such as mean temperature). Another important direction is to learn likelihood-calibrated predictive distributions that enable rigorous probabilistic scoring and calibration analysis, and to formulate and evaluate emulators for cross-model generalization, in line with emerging multi-model benchmarking datasets.

\section{Background}
\paragraph{Neural SDE:} A stochastic differential equation (SDE) models continuous-time evolution with random forcing:
\begin{equation}
\label{s}
\mathrm{d}x_t = f(x_t,t)\,\mathrm{d}t + g(x_t,t)\,\mathrm{d}W_t
\end{equation}
where $f$ is the drift, $g$ the diffusion, and $W_t$ a Wiener process better known as Brownian motion. The Brownian term injects state/time-dependent stochastic forcing via $g(x_t,t)$. In Ito form, the state density $p(x,t)$ obeys the forward Kolmogorov (Fokker–Planck) equation describing the time-evolution of the PDE and the related Stratonovich form corresponding the classical chain rule and is coordinate-invariant, which is often advantageous in physics.

Neural SDE introduced by \cite{kidger2021neural} replaces $f$ and $g$ with neural networks $f_\theta$ and $g_\theta$ \cite{shen2025neural}. The result is a learned stochastic flow that is continuous in time and probabilistic, which naturally handles irregular sampling, missing data, variable horizons and encodes heteroscedastic aleatoric uncertainty. Training a neural SDE on negative log-likelihood objective with re-parameterized Brownian paths gives calibrated and input-dependent uncertainty. 

Wiener (Brownian) process is a continuous path with independent Gaussian increments, so over a step $\Delta t$, the increment satisfies $\Delta W_t \sim \mathcal N(0,\Delta t)$, where $\mathcal N$  is normal (Gaussian) distribution with mean 0 and variance $\Delta t$. In the SDE, $\mathrm{d}W_t$ provides white-noise forcing representing fast, unresolved processes. In discrete time the increment is sampled as $\Delta W_t \approx \sqrt{\Delta t}\,\varepsilon$ with $\varepsilon\sim\mathcal N(0,I)$ with standard multivariate normal and identity covariance $I$. In  \eqref{s}, the drift network $f_\theta$ sets the mean tendency, while the diffusion $g_\phi(x,t)$ scales this stochastic input state-dependently. 

Sampling different Brownian sequences $\{\varepsilon_k\}$ yields different trajectories from the same initial condition, forming an ensemble, aggregating many such realizations characterizes the predictive distribution and when used in NLL training, reduces gradient variance by averaging over noise paths.

\section{Methodology}
\label{sec:method}

\begin{figure*}
    \centering
    \includegraphics[scale=0.45]{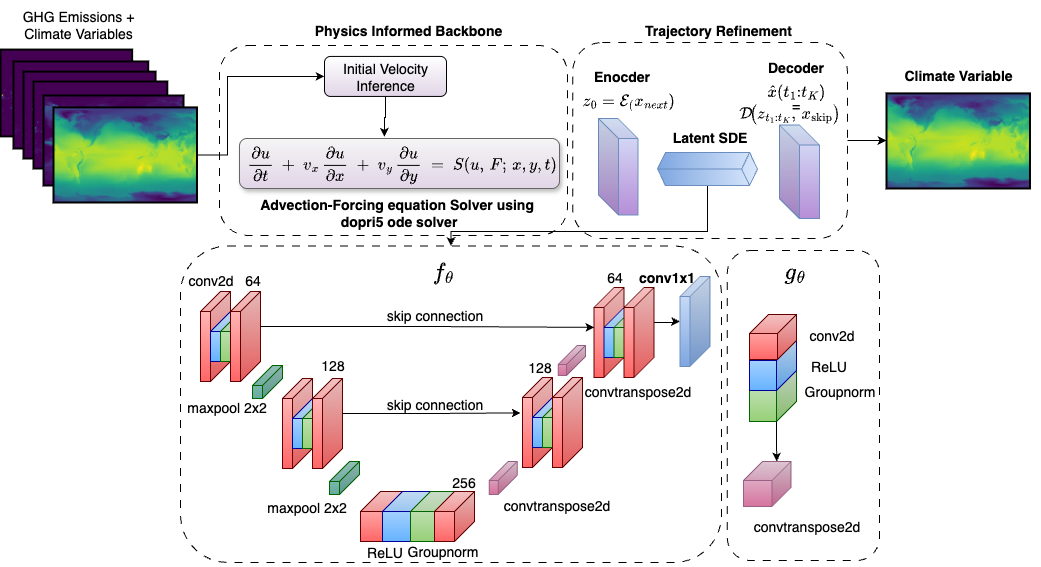}
\caption{Architectural pipeline of A-UTE. Forcing maps and the initial
temperature field are used as input for the physics-informed backbone (left), where an initial pseudo-velocity is inferred and the forced-advection equation is integrated with the \texttt{dopri5} ODE solver to produce a transport-shaped backbone trajectory. This trajectory is refined by the latent neural SDE (right): an encoder $\mathcal{E}_\theta$ maps it to a latent state, an It\^o SDE with drift $f_\theta$ and diffusion $g_\theta$ evolves the latent trajectory, and a decoder $\mathcal{D}_\theta$ returns the refined temperature field. The full pipeline is trained end-to-end under the latitude-weighted Gaussian negative log-likelihood objective.}
\label{fig:arch}
\end{figure*}

\paragraph{Notation.}
$T_t \in \mathbb{R}^{C \times H \times W}$ is the monthly-mean near-surface air temperature (TAS) field at month $t$ on an $H \times W$ latitude-longitude grid ($C$ temperature channels), and $f_t \in \mathbb{R}^{F \times H \times W}$ is the stack of $F$ prescribed monthly forcing maps. The model advances in steps of $\Delta t = 1$ month. $\Phi_{\Delta t}$ denotes the advection-forcing ODE core and $R_\theta$ the learned residual; $\mathbf{c}=(c_x,c_y)$ is the pseudo-velocity field. The latent residual uses an encoder $\mathcal{E}_\theta$ and decoder $\mathcal{D}_\theta$, with latent state $z$ evolving under drift $f_\theta$ and diffusion $g_\theta$. We draw $K$ Brownian samples per window, summarised by a predictive mean $\mu$ and variance $\sigma^2$.

\subsection{Problem Formulation}
\label{sec:formulation}
We formulate temperature emulation as a probabilistic, continuous-time,
auto-regressive prediction task. Given an initial state $T_{t_0}$ and a
prescribed forcing sequence $f_{t_0:t_H}$, the objective is to learn the conditional predictive distribution $P_\theta(T_{t_1:t_H} \mid T_{t_0}, f_{t_0:t_H})$ and to generate stable decade-scale roll-outs. A-UTE is built as a two-stage surrogate. An \emph{advection-forcing dynamical core} integrates an advection-with-forcing equation over $[t, t+\Delta t]$ to produce an intermediate trajectory that acts as a transport-consistent prior for the coarse-grained spatial evolution of monthly-mean TAS \cite{adem1970incorporation}. A \emph{neural SDE residual} then learns a stochastic correction in a latent space, representing the unresolved processes and aggregation effects that monthly averaging and coarse resolution cannot explicitly resolve. 


The one-step operator composes the two stages: the ODE core maps the current state and forcing to a deterministic backbone trajectory, and the neural-SDE pathway refines that trajectory into the next field,
\begin{equation}
  \begin{aligned}
    \tilde{T}_{t:t+\Delta t} &= \Phi_{\Delta t}\!\left(T_t, f_{t:t+\Delta t}\right), \\
    T_{t+\Delta t}
      &= \mathcal{R}_\theta\!\left(\tilde{T}_{t:t+\Delta t}\right)
       = \big(\mathcal{D}_\theta \circ \mathrm{SDE}_\theta \circ \mathcal{E}_\theta\big)\!\left(\tilde{T}_{t:t+\Delta t}\right).
  \end{aligned}
  \label{eq:onestep}
\end{equation}
where $\Phi_{\Delta t}$ integrates the advection-forcing ODE (Sec.~\ref{sec:ode}) to produce the deterministic backbone $\tilde{T}$, and $\mathcal{R}_\theta$ denotes the neural-SDE refinement (Sec.~\ref{sec:sde}), i.e. the encoder-SDE-decoder pathway of Eqs.~\eqref{eq:encoder}-\eqref{eq:decoder}. The forcing $f$ enters $\mathcal{M}_\theta$ only through the source term $S(T,f)$ of the ODE core; the encoder, latent SDE, and decoder act on the backbone trajectory $\tilde{T}$ and therefore depend on the forcing solely through it. Repeated application of \eqref{eq:onestep} yields an auto-regressive roll-out from $T_{t_0}$ under $f_{t_0:t_H}$.

\subsection{Advection-Forcing Dynamical Core (ODE)}
\label{sec:ode}
A-UTE uses a forced-advection dynamical core as a structured prior for the evolution of monthly-mean TAS on a latitude-longitude grid. At monthly resolution the target field is a coarse-grained thermodynamic state in which fast variability is averaged out, while large-scale transport remains a key mechanism shaping spatial temperature patterns over time. We model the deterministic backbone over each integration window with a two-dimensional forced-advection equation,
\begin{equation}
  \frac{\partial T}{\partial t}
  + c_x \frac{\partial T}{\partial x}
  + c_y \frac{\partial T}{\partial y}
  = S\!\left(T, f;\, x, y, t\right),
  \label{eq:pde}
\end{equation}
where $T(x,y,t)$ denotes monthly-mean TAS, $\mathbf{c}(x,y,t)=(c_x,c_y)$ is an effective horizontal pseudo-velocity field on the grid, and $S(\cdot)$ is a forcing term parameterised from the prescribed external drivers $f$ over the same interval. We emphasise that $\mathbf{c}$ is not an observed wind field but an effective transport field derived from the temperature data itself, so \eqref{eq:pde} acts as a transport-shaped prior rather than a closed physical representation of the surface energy budget.

\paragraph{Numerical discretisation and integration.}
We solve \eqref{eq:pde} as a semi-discrete system, approximating spatial derivatives with centred finite differences on the regular grid \cite{fiadeiro1977weighted, molenkamp1968accuracy, leveque2007finite}. For grid indices $(i,j)$, with $i$ the latitude row and $j$ the longitude column, and time level $t^n$,
\begin{equation}
  \left(\frac{\partial T}{\partial x}\right)_{i,j}
  \approx \frac{T^{\,n}_{i,j+1} - T^{\,n}_{i,j-1}}{2\,\Delta x},
  \qquad
  \left(\frac{\partial T}{\partial y}\right)_{i,j}
  \approx \frac{T^{\,n}_{i+1,j} - T^{\,n}_{i-1,j}}{2\,\Delta y}.
  \label{eq:fd}
\end{equation}
This yields an ODE in time (with spatially discretised tendencies), which we integrate over each window using the Dormand-Prince method (\texttt{dopri5}) \cite{dormand1980family}.

\paragraph{Boundary conditions.}
The centred finite-difference operators require boundary handling that respects the grid topology (periodic in longitude, non-periodic in latitude) to avoid artificial discontinuities. Longitude is treated as cyclic, so we impose wrap-around indexing
\begin{equation}
  T_{i,0} \equiv T_{i,W}, \qquad T_{i,W+1} \equiv T_{i,1},
  \label{eq:bc-lon}
\end{equation}
and apply \eqref{eq:fd} directly. Latitude is not periodic, so we use replicate padding across the first and last rows,
\begin{equation}
  T_{0,j} \equiv T_{1,j}, \qquad T_{H+1,j} \equiv T_{H,j},
  \label{eq:bc-lat}
\end{equation}
which keeps the centred-difference operator well-defined at the boundary rows while preventing cross-pole periodicity.

\paragraph{Effective transport field: initialisation and refinement.}
To advect monthly-mean TAS within the dynamical core, we construct the pseudo-velocity field from the local spatiotemporal evolution of the temperature field. At the start of each roll-out window $t_0$, we estimate the local temporal tendency $\partial_t T(x,y,t_0)$ from a short history using a smooth temporal fit at each grid cell (a natural cubic spline) \cite{mckinley1998cubic}, and compute the spatial gradient $\nabla T(x,y,t_0)$ by finite differences. We then define a dimensionally consistent migration-speed proxy
\begin{equation}
  \lVert \mathbf{c} \rVert (x,y)
  = \frac{\lvert \partial_t T(x,y,t_0) \rvert}
         {\lVert \nabla T(x,y,t_0) \rVert + \varepsilon},
  \label{eq:speed}
\end{equation}
where $\varepsilon > 0$ ensures numerical stability when $\lVert \nabla T \rVert$ is near zero. Equation~\eqref{eq:speed} has units of velocity: the numerator measures the local rate of temperature change, and the denominator converts that change to an equivalent spatial displacement rate via the local gradient magnitude. We obtain an initial vector field by pairing $\lVert \mathbf{c} \rVert (x,y)$ with a local direction proxy aligned with $\nabla T$ and a sign consistent with the advection convention, and subsequently refine the resulting transport field during ODE integration with a lightweight learned correction module based on a small $1\times1$ convolutional MLP trained end-to-end under the downstream likelihood objective.

\subsection{Neural SDE Residual (Probabilistic Refinement)}
\label{sec:sde}
The forced-advection dynamical core provides a deterministic, physics prior backbone trajectory over each roll-out window. At monthly resolution, the observed TAS evolution also reflects aggregated sub-monthly variability and systematic discrepancy not represented by the coarse-grained core. We therefore augment the deterministic backbone with a probabilistic residual modelled in a compact latent space, which preserves spatial structure via convolutional parameterisations and supports likelihood-based training for calibrated predictive uncertainty.

\paragraph{Window encoding.}
Let $\tilde{T}_{t:t+\Delta}$ denote the intermediate TAS trajectory produced by the ODE core over a window of length $\Delta$. A spatiotemporal encoder $\mathcal{E}_\theta$ maps this window trajectory to an initial latent state,
\begin{equation}
  z_0 = \mathcal{E}_\theta\!\left(\tilde{T}_{t:t+\Delta}\right),
  \label{eq:encoder}
\end{equation}
where $z_0$ is a latent tensor on a reduced grid.

\paragraph{Latent stochastic dynamics.}
Starting from $z_0$, the latent state is evolved with an It\^o SDE with
diagonal noise,
\begin{equation}
  \mathrm{d}z_s = f_\theta(z_s, s)\,\mathrm{d}s + g_\theta(z_s, s)\,\mathrm{d}W_s,
  \qquad z_{s_0} = z_0,
  \label{eq:latentsde}
\end{equation}

where $s$ is the continuous integration variable within the window and $W_s$ is a standard Wiener process. The learnable convolutional maps $f_\theta$ and $g_\theta$ parameterise the latent SDE. $f_\theta$ specifies the deterministic residual tendency (drift), while $g_\theta$ specifies the state-dependent noise amplitude (diffusion). Stochasticity enters through the increments
$\mathrm{d}W_s$, sampling independent Brownian paths yields multiple latent trajectories and, after decoding, an ensemble of refined roll-outs.

\paragraph{Decoding to physical space.}
The latent trajectory is mapped back to the physical grid by a decoder
$\mathcal{D}_\theta$,
\begin{equation}
      \hat{T}_{t:t+\Delta} = \mathcal{D}_\theta\!\left(z_{s_0:s_K};\, \tilde{T}_{t:t+\Delta}\right),
\label{eq:decoder}
\end{equation}
producing a refined TAS trajectory over the same window. The backbone trajectory $\tilde{T}$ re-enters through a skip connection (Fig.~\ref{fig:arch}), so the pathway refines rather than replaces the transport prior. The encoder-SDE-decoder pathway is trained end-to-end so that the stochastic residual corrects both unresolved variability and systematic mismatch relative to the target monthly TAS evolution.

\paragraph{Predictive distribution and roll-out semantics.}
For each window we draw $K$ independent Brownian paths $ \{W^{(k)}\}_{k=1}^{K}$ and integrate \eqref{eq:latentsde} to obtain $K$ refined trajectories $\{\hat{T}^{(k)}_{t:t+\Delta}\}_{k=1}^{K}$. We summarise these samples with a window-level mean and variance,
\begin{equation}
  \mu = \frac{1}{K}\sum_{k=1}^{K} \hat{T}^{(k)},
  \qquad
  \sigma^2 = \frac{1}{K-1}\sum_{k=1}^{K}\left(\hat{T}^{(k)} - \mu\right)^2,
  \label{eq:meanvar}
\end{equation}
computed pointwise on the latitude-longitude grid. The auto-regressive state passed to the next window is the predictive mean $\mu$, while $\sigma^2$ provides the within-window predictive spread used for probabilistic evaluation and uncertainty reporting.

\subsection{Uncertainty Quantification}
\label{sec:objective}
Existing climate emulators, including probabilistic ones, are commonly trained with a deterministic objective such as MSE/RMSE, which fits the conditional mean but leaves the predictive spread under-determined. To learn both the mean and a heteroscedastic variance, we train with a latitude-weighted Gaussian negative log-likelihood (NLL) objective. Consider a dataset $\mathcal{D} = \{(t_i, y_i)\}_{i=1}^{N}$, where each frame $y_i \in \mathbb{R}^{C \times H \times W}$ lies on a regular latitude-longitude grid. At time $t_i$ the model predicts a cell-wise mean $\mu_\theta(t_i)$ and variance $\sigma^2_\theta(t_i)$. The latitude-weighted Gaussian NLL with a
variance prior is
\begin{equation}
\label{eq:nll}
\begin{aligned}
\mathcal{NLL}
&=
-\frac{1}{NCHW}
\sum_{i=1}^{N}
\Bigg[
\sum_{p=1}^{CHW}
\tilde w_{p}\;
\log \mathcal{N}\!\big(y_{i,p}\,\big|\,\mu_{\theta,i,p},\,\sigma^{2}_{\theta,i,p}\big)
\\
&\qquad\qquad\qquad\qquad
+\;
\log \mathcal{N}^{+}\!\big(\sigma^{2}_{\theta,i}\,\big|\,0,\,\lambda_\sigma^{2}\mathbf{I}\big)
\Bigg].
\end{aligned}
\end{equation}
where the flattened index $p = 1,\dots,CHW$ runs over channel $c$, latitude row $\ell$, and longitude column $m$. $y_{i,p}$, $\mu_{\theta,i,p}$, and $\sigma^2_{\theta,i,p}$ are the observed value, predicted mean, and predicted variance at a single space-channel location of frame $i$. $\tilde{w}_p$ is the cosine-of-latitude weight for location $p$, $\log\mathcal{N}$ is the Gaussian log-likelihood, and $\log\mathcal{N}^{+}$ is a half-Gaussian prior over the non-negative variance vector of frame $i$ with scale $\lambda_\sigma$, which shrinks extreme variances and stabilises training.

\section{Experiments and Results}
\label{sec:results}

\subsection{Datasets and Task}
\label{sec:data}

\paragraph{Task.}
A-UTE auto-regressively emulates the monthly temperature state (TAS) from an initial condition and a parallel time series of prescribed forcings. We train with 1-year auto-regressive roll-outs and evaluate on 10-year (120-step) roll-outs, and we assess cross-model generalisation through zero-shot emulation. Training hyperparameters of A-UTE are given in Table \ref{hyper-aiet} and \ref{hyper-aiet1}. We compare A-UTE against the U-Net and ConvLSTM emulators distributed with ClimateSet \cite{kaltenborn2023climateset}, the SFNO configuration from LUCIE \cite{guan2024lucie, bonev2023spherical}, and NADE \cite{choi2023climate} adapted to gridded TAS. Our code is available at \url{https://github.com/cruiseresearchgroup/A-UTE}.

\begin{table}[H]
\caption{A-UTE Training Details}
  \centering
    \label{hyper-aiet}
     \begin{tabular}{ll}
\multicolumn{1}{l}{\bf Hyperparameters} &\multicolumn{1}{l}{\bf Value} \\
\toprule
    Epochs & 30 \\
    Conv2d kernel size & 3x3 \\ 
    Integrator & srk \\
    Noise type & diagnol \\
    ODE solver & dopri5 \\
    Pooling No.1 & AvgPool2d (2,2) \\
    Normalization & GroupNorm \\
    Activation Function & ReLU \\
    Optimizer & Adam \\
    Learning Rate & $1e-4$ \\
    Weight Decay & $1e-6$ \\
    Adam $\epsilon$ & $1e-8$ \\
    LR scheduler & Exponential decay \\
    Scheduler gamma & $0.98$ \\
    Batch size & 4
\\ \bottomrule
\end{tabular}
\end{table}

\begin{table}[H]
\caption{A-UTE ConvLayer Details}
    \centering
    \label{hyper-aiet1}
    \begin{tabular}{ll}
    \multicolumn{1}{l}{\bf ConvLayers} &\multicolumn{1}{l}{\bf Value}\\ 
    \toprule
    Encoder & 4 \\
    Latent SDE - Drift $f_\theta$ & 11 \\ 
    Latent SDE - Diffusion $g_\theta$& 1 \\ 
    Decoder & 4 \\
    Velocity Refiner & 2 \\
    \hline
    \end{tabular}
\end{table} 

\paragraph{ClimateSet.}
We train A-UTE on 20 physics-based climate models: 19 distributed with ClimateSet \cite{kaltenborn2023climateset} and FV3GFS \cite{zhou2019toward}. ClimateSet compiles outputs from the Coupled Model Intercomparison Project Phase~6 (CMIP6) \cite{eyring2016overview}, drawing model responses from ScenarioMIP \cite{o2016scenario} and forcing trajectories from Input4MIPs \cite{durack2017input4mips}. Each model is standardised to a common $96\times144$ ($\approx$250\,km) latitude-longitude grid at monthly temporal resolution, providing 86-year series (2015-2100) across the four core SSP scenarios (SSP1-2.6, SSP2-4.5, SSP3-7.0, SSP5-8.5). We train on SSP1-2.6, validate on SSP3-7.0, and test on the held-out SSP2-4.5 scenario. This cross-scenario split probes forcing levels and responses outside the training range. The target is monthly-mean near-surface air temperature (TAS), and the forcing inputs are CO\textsubscript{2}, CH\textsubscript{4}, SO\textsubscript{2}, and black carbon.

\paragraph{FV3GFS.}
We also train A-UTE on FV3GFS \cite{zhou2019toward}, the operational global forecast system of the US National Weather Service and the National Centers for Environmental Prediction. It comprises an 11-member initial-condition ensemble, each a 10-year integration saved every 6 hours, with forcings of annually repeating climatological sea-surface temperature (1982-2012 mean) and top-of-atmosphere insolation. Output is conservatively re-gridded from the FV3 cubed-sphere to a $180\times360$ ($1^\circ$) Gaussian grid and passed through a spherical-harmonic analysis-synthesis to suppress high-latitude artefacts. We downsample to a monthly cadence to align with ClimateSet, train on 100 years drawn from 10 ensemble members, and evaluate on a distinct 10-year member. The target is air temperature at multiple vertical levels.

The 10-year (120-step) horizon follows the evaluation data: FV3GFS provides 10-year integrations, and the ClimateSet 2015-2100 series is partitioned so that testing evaluates the held-out SSP2-4.5 scenario outside the training forcing range. This span spans both the recurring seasonal cycle and the long-term forced trend of monthly-mean TAS. The seasonal cycle is retained at every step of the roll-out, explicitly through the downward shortwave flux for FV3GFS and implicitly through the temperature field for ClimateSet, while the forced trend is driven by the time-varying forcing maps and tested directly by the cross-scenario split. The bounded RMSE across the full 120-step roll-out (Section~\ref{sec:forced}) indicates that both are tracked without the runaway error growth reported for some data-driven emulators at multi-year horizons.

\subsection{Baselines and Metrics}
\label{sec:metrics}

\paragraph{Baselines.}
U-Net and ConvLSTM follow the ClimateSet configurations, adapted from the diagnostic to the auto-regressive setting. SFNO follows the LUCIE configuration, and NADE is adapted to the $96\times144$ grid with monthly TAS as the target. NADE is the closest baseline, sharing an explicit transport-shaped PDE structure inside a learned dynamical model.

\paragraph{Evaluation Metrics.}
We report latitude-weighted root-mean-square error (RMSE) (Equation \ref{eq:rmse}) \cite{nguyen2023climax} for deterministic accuracy and the continuous ranked probability score (CRPS) \cite{winkler1996scoring} for probabilistic skill (Equation \ref{eq:crps_gauss_single}. A-UTE yields a Gaussian predictive distribution directly from the $K$ Brownian realisations of the latent SDE. The four baselines are deterministic and to enable a CRPS comparison, we construct an empirical predictive distribution for each by training and evaluating it five times with different random seeds and treating the five roll-outs as samples to produce their probabilistic score.

\begin{equation}
\label{eq:rmse}
\mathrm{RMSE}
=
\frac{1}{N}\sum_{t=1}^{N}
\sqrt{
\frac{1}{HW}
\sum_{h=1}^{H}\sum_{w=1}^{W}
L(h)\,\big(y_{t,h,w}-\hat{y}_{t,h,w}\big)^{2}
},
\end{equation}
where $L(h)$ are latitude weights (constant over $w$ at a fixed latitude index $h$).

\begin{equation}
\label{eq:latw}
L(h)
=
\frac{\cos(\mathrm{lat}(h))}
{\frac{1}{H}\sum_{h'=1}^{H}\cos(\mathrm{lat}(h'))}.
\end{equation}

\begin{equation}
\label{eq:crps_gauss_single}
\begin{aligned}
\mathrm{CRPS}\!\left(\mathcal{N}(\mu,\sigma^{2}),\,y\right)
&=
\sigma\!\left[
z\!\left(2\Phi(z)-1\right)
+2\phi(z)-\frac{1}{\sqrt{\pi}}
\right],\\
z&=\frac{y-\mu}{\sigma}.
\end{aligned}
\end{equation}

\subsection{Forced-Response Results}
\label{sec:forced}

Table~\ref{tab:climateset} summarises 10-year auto-regressive roll-outs across the 19 ClimateSet models under the held-out SSP2-4.5 scenario. A-UTE attains the lowest RMSE and CRPS on every model. Its RMSE ranges from 0.333\,K (CESM2-WACCM) to 0.420\,K (CAS-ESM2-0) with a median of 0.372\,K. The strongest deterministic baseline, SFNO, has a median near 0.418\,K, while ConvLSTM and U-Net cluster near 0.55\,K and 1.0\,K respectively. The advantage holds across modelling centres, atmospheric resolutions, and structural model differences, and is obtained under the cross-scenario split, so the consistency of the lead indicates that the emulator generalises beyond the training forcing range rather than overfitting it. A-UTE's aggregated RMSE remains bounded across the full 120-step roll-out, with every model inside the 0.33-0.42\,K range, consistent with the absence of the runaway error growth reported for some purely data-driven roll-out emulators at multi-year horizons
\cite{chattopadhyay2023long}.

The separation of roles underlies these gains: the advection prior supplies a transport-consistent inductive bias for the large-scale spatial redistribution of the monthly-mean temperature field, while the latent neural SDE absorbs the systematic discrepancy between that prior and the target model evolution and exposes a calibrated predictive spread. Against NADE, which embeds an advection-diffusion structure on a graph with a deterministic objective on regional data, A-UTE applies a forced-transport core coupled with a stochastic latent residual on global grids, trained end-to-end under a Gaussian log-likelihood.

\begin{table*}[!h]
\centering
\caption{10-year auto-regressive roll-outs on 19 ClimateSet models under the held-out SSP2-4.5 scenario. Latitude-weighted RMSE (K) and CRPS (K); lower is better. CRPS for the deterministic baselines is computed from a 5-seed empirical ensemble (Section~\ref{sec:metrics}). Best entry per metric in bold.}
\label{tab:climateset}
\small
\begin{tabular}{l cc cc cc cc cc}
\toprule
& \multicolumn{2}{c}{A-UTE} & \multicolumn{2}{c}{U-Net} & \multicolumn{2}{c}{ConvLSTM} & \multicolumn{2}{c}{SFNO} & \multicolumn{2}{c}{NADE} \\
\cmidrule(lr){2-3}\cmidrule(lr){4-5}\cmidrule(lr){6-7}\cmidrule(lr){8-9}\cmidrule(lr){10-11}
Climate Model & RMSE & CRPS & RMSE & CRPS & RMSE & CRPS & RMSE & CRPS & RMSE & CRPS \\
\midrule
AWI-CM-1-1-MR    & \textbf{0.369} & \textbf{0.251} & 0.916 & 0.842 & 0.543 & 0.441 & 0.406 & 0.316 & 0.412 & 0.327 \\
BCC-CSM2-MR      & \textbf{0.381} & \textbf{0.263} & 0.950 & 0.871 & 0.549 & 0.451 & 0.412 & 0.323 & 0.435 & 0.355 \\
CAS-ESM2-0       & \textbf{0.420} & \textbf{0.306} & 1.431 & 0.921 & 0.542 & 0.439 & 0.451 & 0.362 & 0.432 & 0.330 \\
CESM2-WACCM      & \textbf{0.333} & \textbf{0.233} & 1.154 & 0.932 & 0.556 & 0.459 & 0.463 & 0.365 & 0.461 & 0.351 \\
CESM2            & \textbf{0.353} & \textbf{0.248} & 0.926 & 0.811 & 0.557 & 0.457 & 0.475 & 0.397 & 0.387 & 0.291 \\
CMCC-ESM2        & \textbf{0.387} & \textbf{0.266} & 1.173 & 0.910 & 0.552 & 0.431 & 0.451 & 0.363 & 0.452 & 0.371 \\
CMCC-CM2-SR5     & \textbf{0.396} & \textbf{0.274} & 0.980 & 0.876 & 0.553 & 0.413 & 0.499 & 0.401 & 0.438 & 0.365 \\
CNRM-CM6-1-HR    & \textbf{0.384} & \textbf{0.266} & 0.943 & 0.850 & 0.543 & 0.403 & 0.411 & 0.323 & 0.418 & 0.333 \\
EC-Earth3        & \textbf{0.372} & \textbf{0.251} & 1.284 & 0.922 & 0.548 & 0.445 & 0.383 & 0.294 & 0.392 & 0.295 \\
EC-Earth3-Veg    & \textbf{0.399} & \textbf{0.273} & 0.987 & 0.870 & 0.549 & 0.445 & 0.418 & 0.329 & 0.408 & 0.314 \\
EC-Earth3-Veg-LR & \textbf{0.405} & \textbf{0.278} & 1.121 & 0.911 & 0.545 & 0.432 & 0.500 & 0.410 & 0.487 & 0.399 \\
FGOALS-f3-L      & \textbf{0.374} & \textbf{0.266} & 1.017 & 0.899 & 0.533 & 0.409 & 0.399 & 0.311 & 0.423 & 0.356 \\
GFDL-ESM4        & \textbf{0.358} & \textbf{0.252} & 0.923 & 0.789 & 0.539 & 0.411 & 0.424 & 0.335 & 0.457 & 0.363 \\
INM-CM4-8        & \textbf{0.348} & \textbf{0.246} & 0.954 & 0.832 & 0.527 & 0.402 & 0.462 & 0.373 & 0.459 & 0.337 \\
INM-CM5-0        & \textbf{0.369} & \textbf{0.261} & 1.128 & 0.924 & 0.529 & 0.404 & 0.394 & 0.315 & 0.412 & 0.313 \\
MPI-ESM1-2-HR    & \textbf{0.352} & \textbf{0.240} & 0.668 & 0.510 & 0.547 & 0.441 & 0.440 & 0.352 & 0.411 & 0.301 \\
MRI-ESM2-0       & \textbf{0.374} & \textbf{0.261} & 0.944 & 0.831 & 0.565 & 0.460 & 0.414 & 0.323 & 0.399 & 0.272 \\
NorESM2-MM       & \textbf{0.363} & \textbf{0.254} & 1.044 & 0.932 & 0.555 & 0.451 & 0.420 & 0.333 & 0.382 & 0.298 \\
TaiESM1          & \textbf{0.353} & \textbf{0.244} & 0.989 & 0.856 & 0.535 & 0.425 & 0.398 & 0.300 & 0.401 & 0.377 \\
\bottomrule
\end{tabular}
\end{table*}

\begin{figure*}
    \centering
    \begin{subfigure}[t]{0.2\textwidth}
        \includegraphics[scale=0.2]{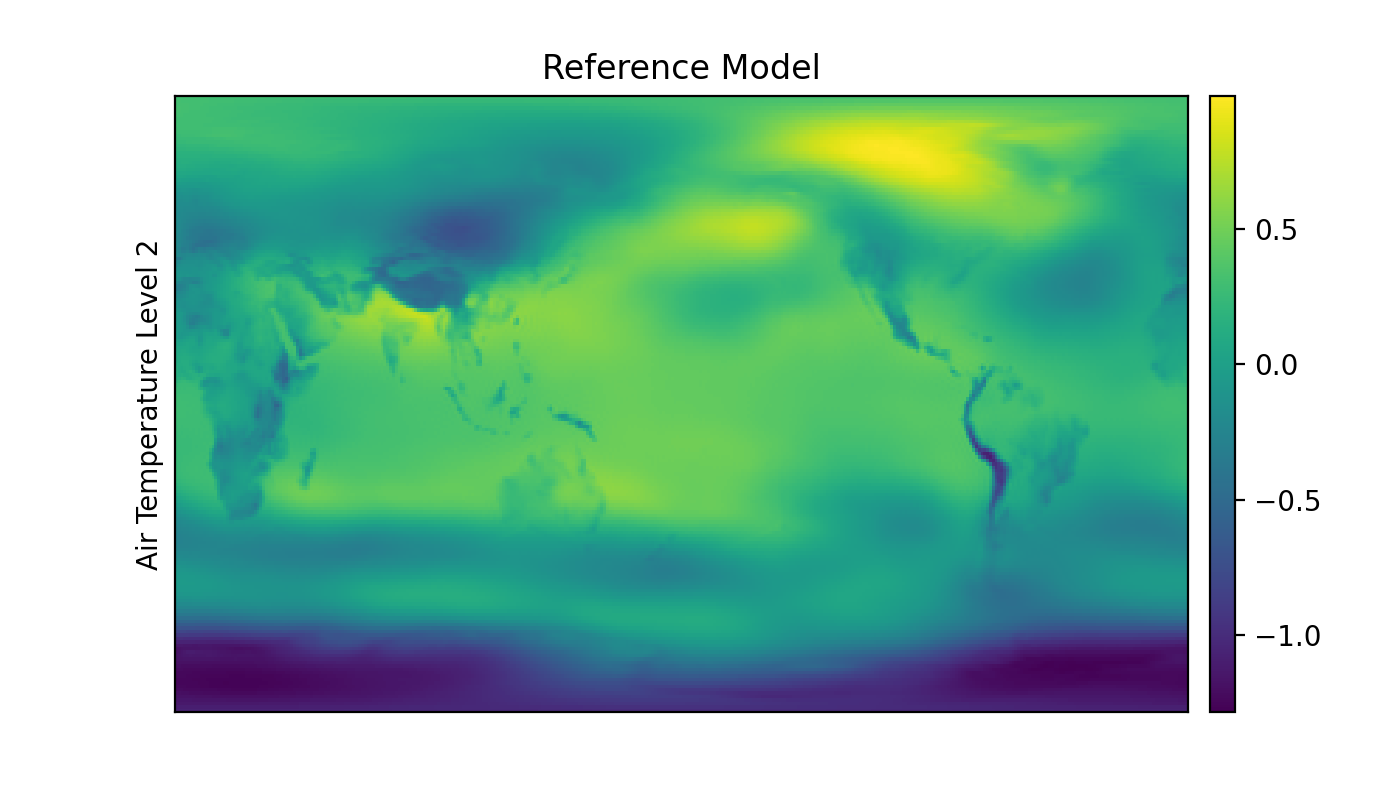}
    \end{subfigure}
    \begin{subfigure}[t]{0.2\textwidth}
        \includegraphics[scale=0.085]{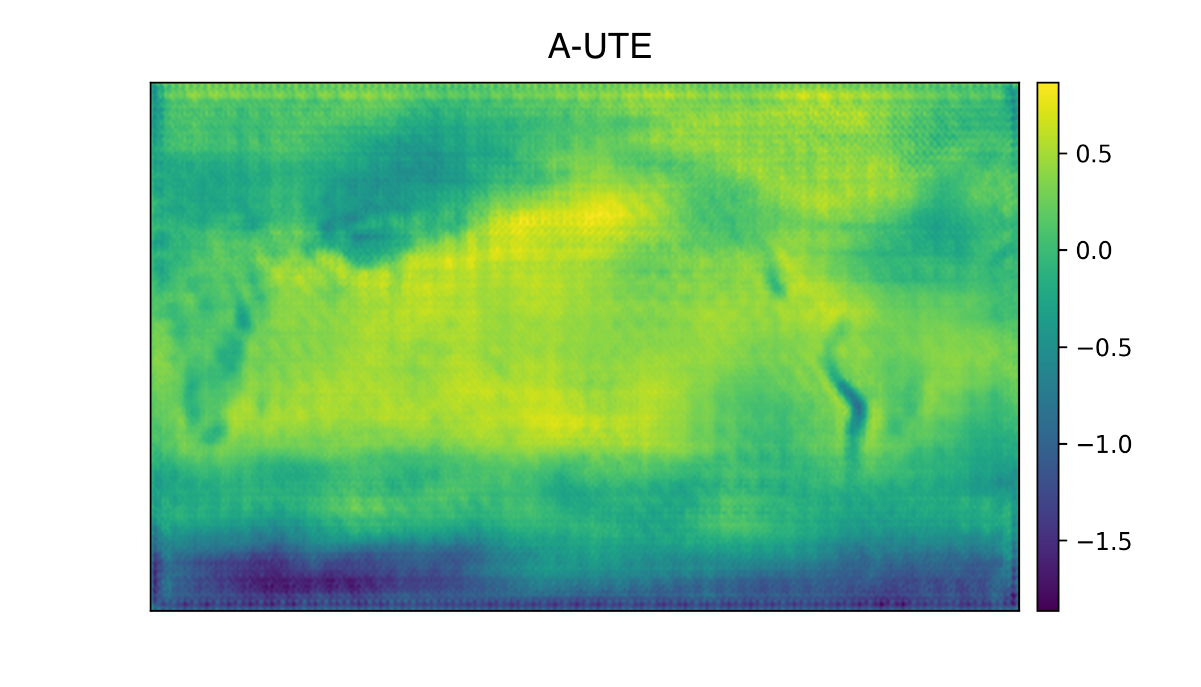}
    \end{subfigure}
    \begin{subfigure}[t]{0.2\textwidth}
        \includegraphics[scale=0.2]{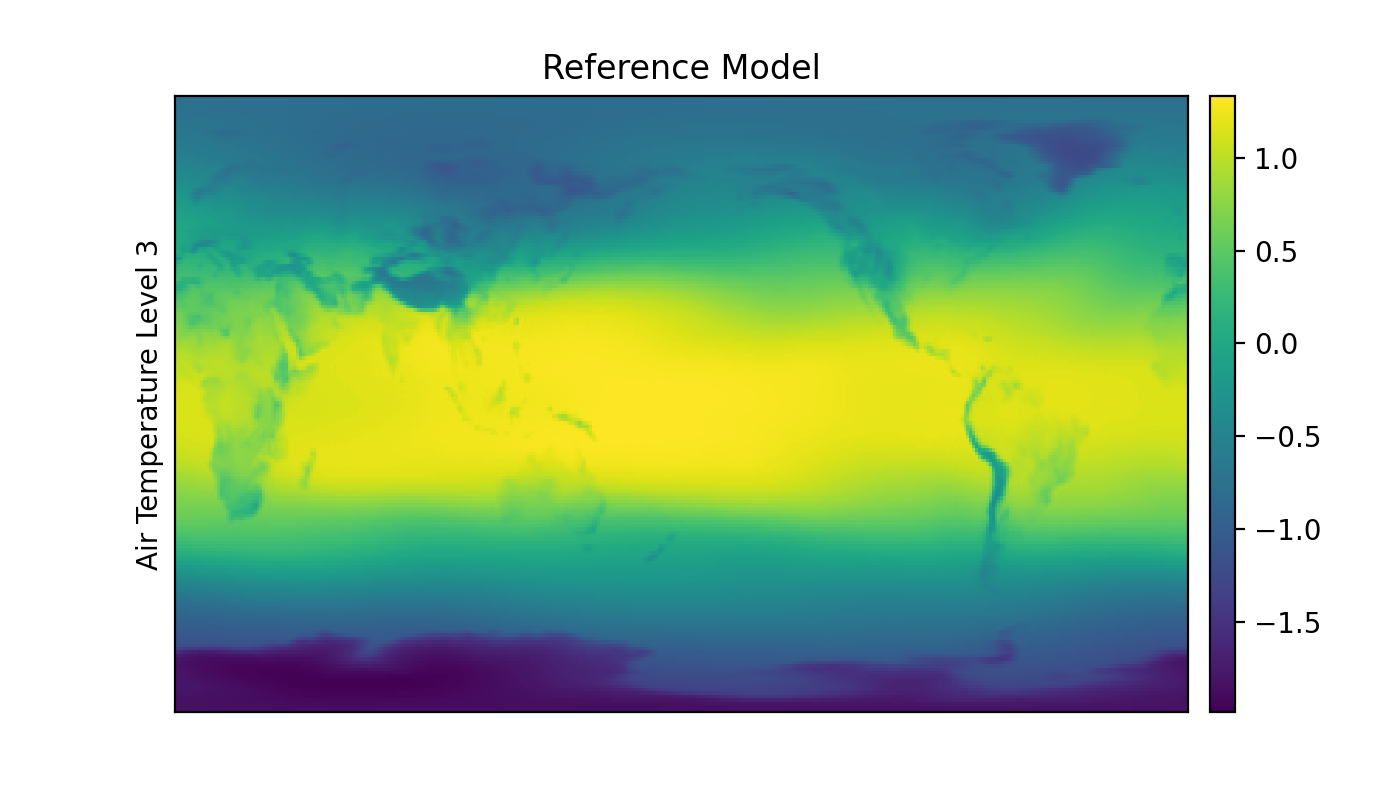}
    \end{subfigure}
    \begin{subfigure}[t]{0.2\textwidth}
        \includegraphics[scale=0.085]{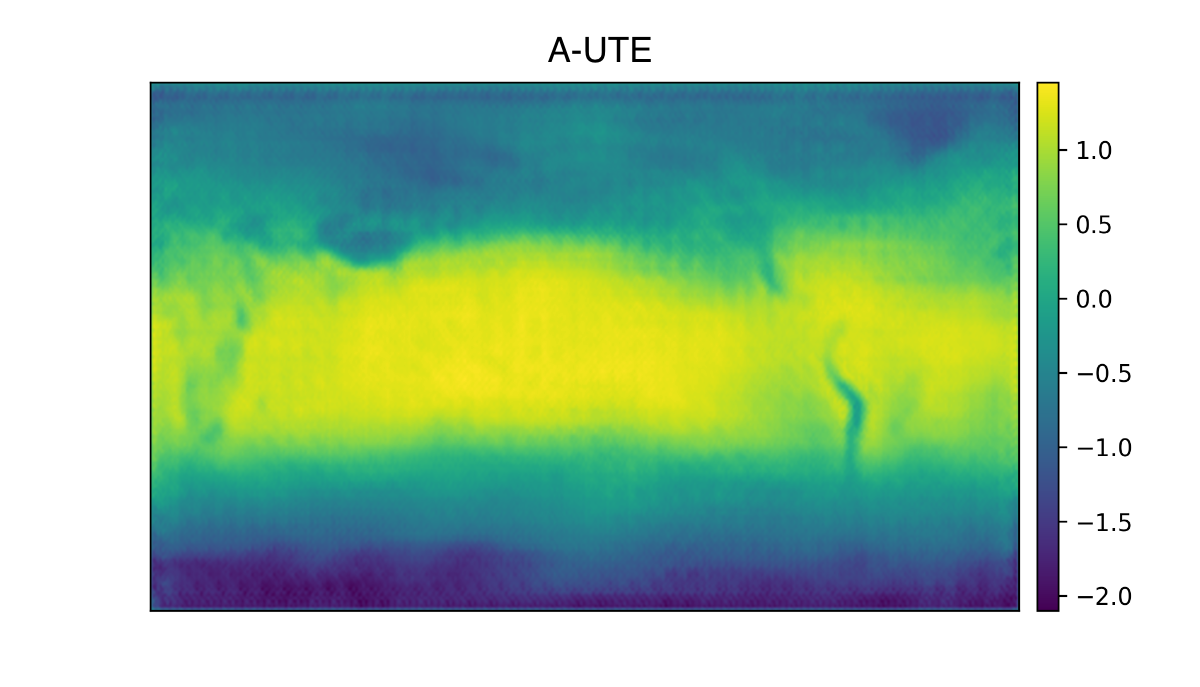}
    \end{subfigure}
    \\
    \begin{subfigure}[t]{0.2\textwidth}
        \includegraphics[scale=0.2]{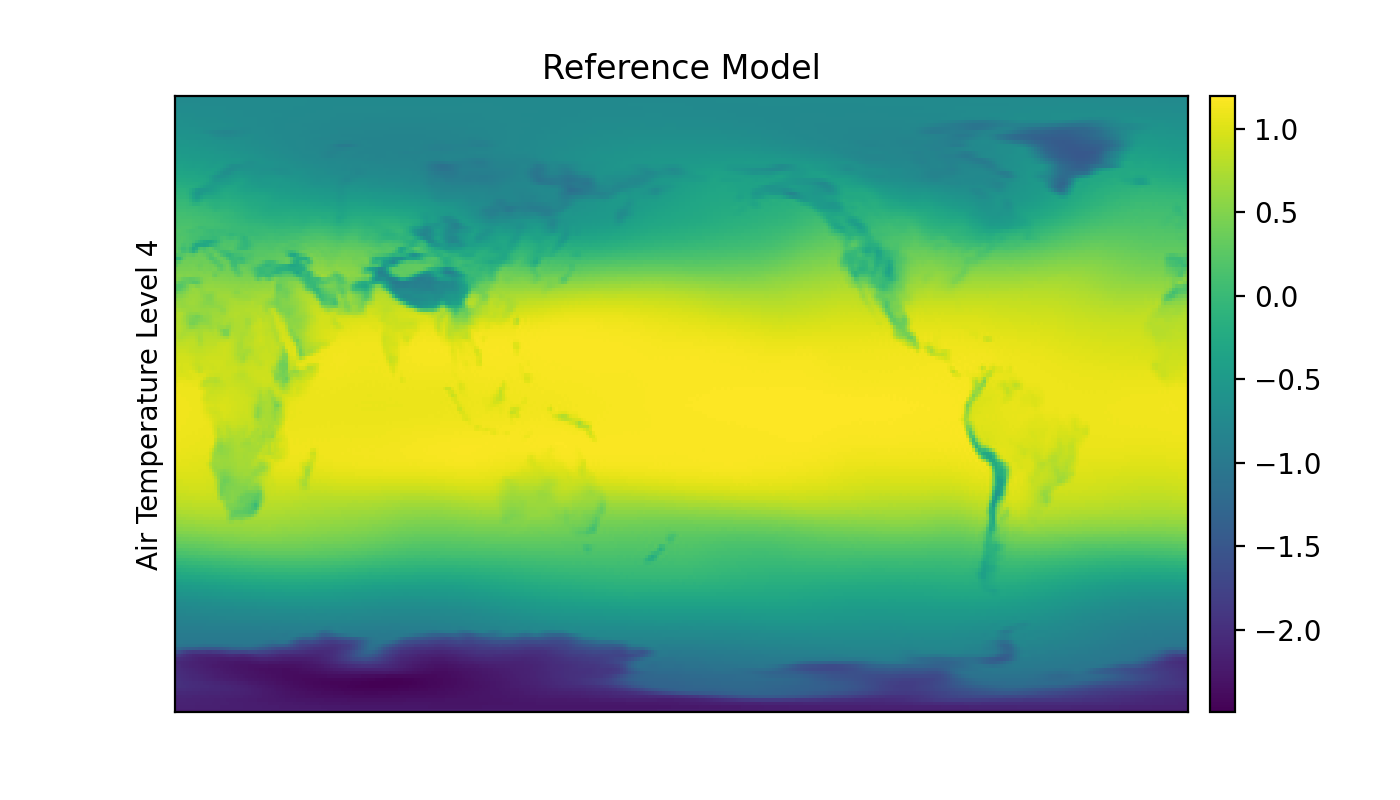}
    \end{subfigure}
    \begin{subfigure}[t]{0.2\textwidth}
        \includegraphics[scale=0.085]{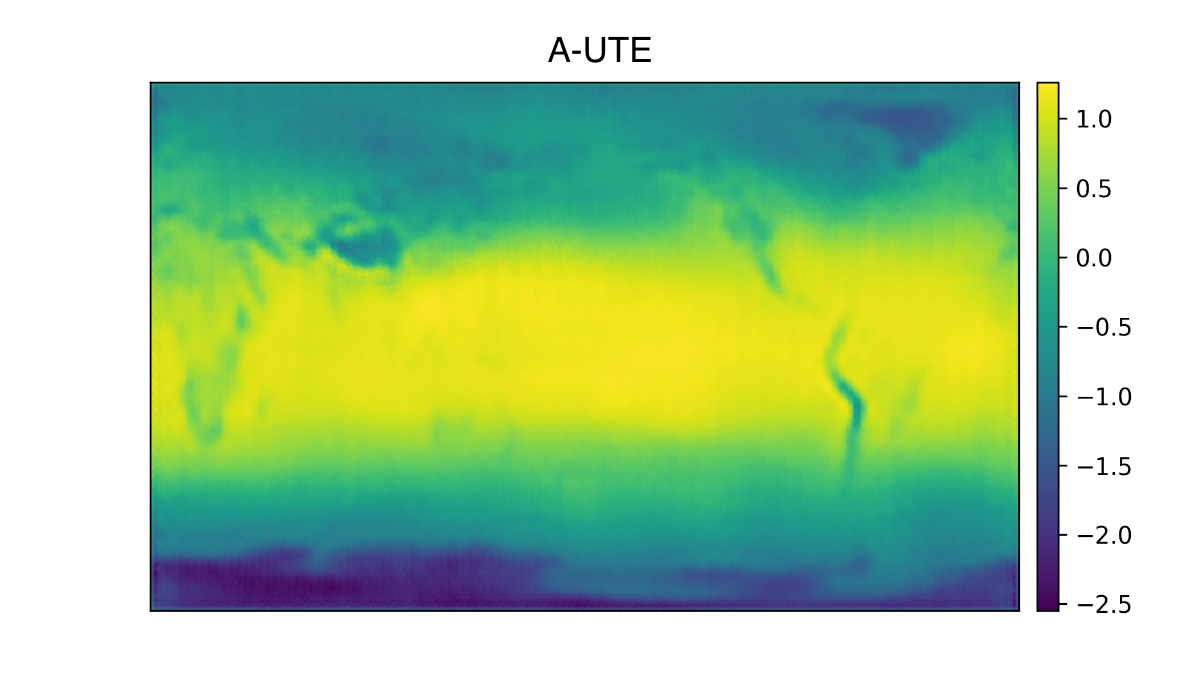}
    \end{subfigure}
    \begin{subfigure}[t]{0.2\textwidth}
        \includegraphics[scale=0.2]{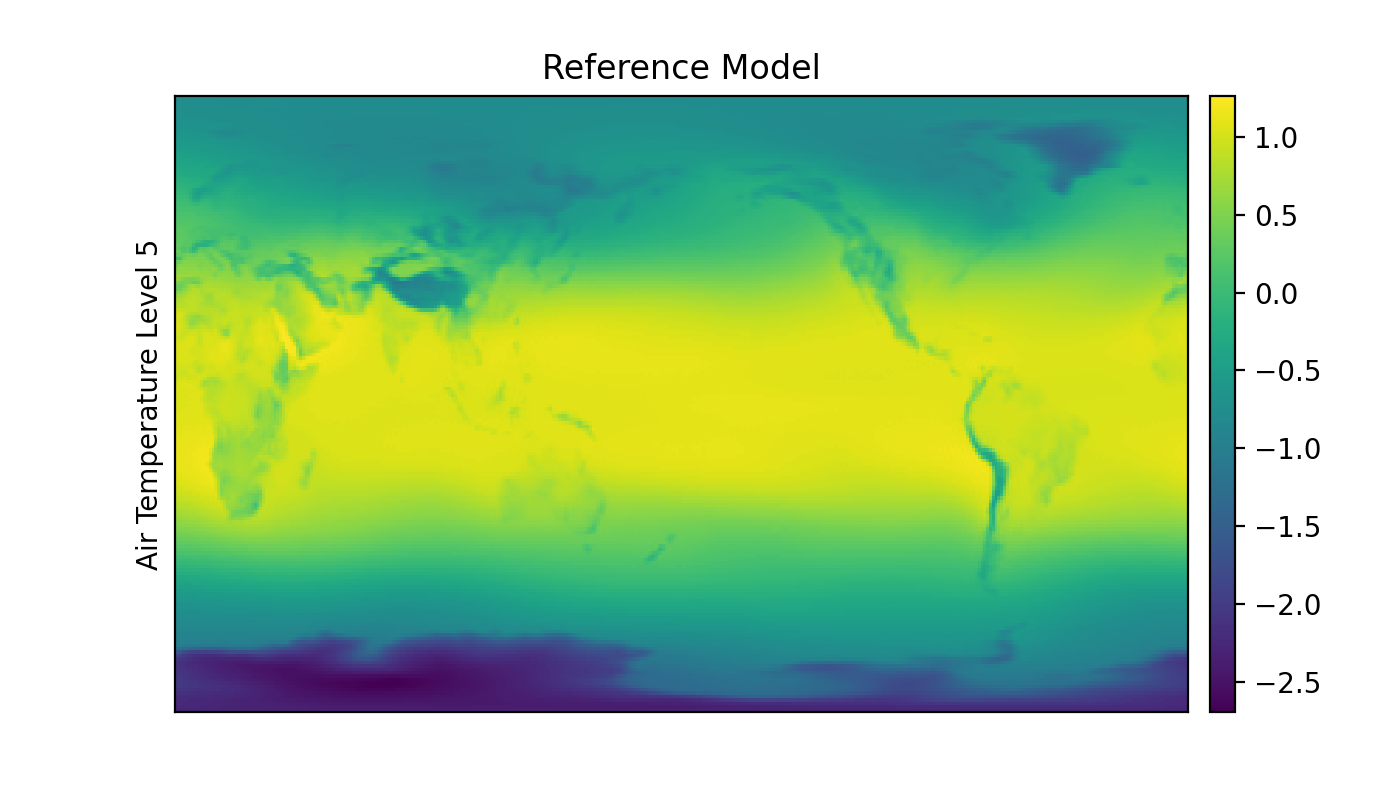}
    \end{subfigure}
    \begin{subfigure}[t]{0.2\textwidth}
        \includegraphics[scale=0.085]{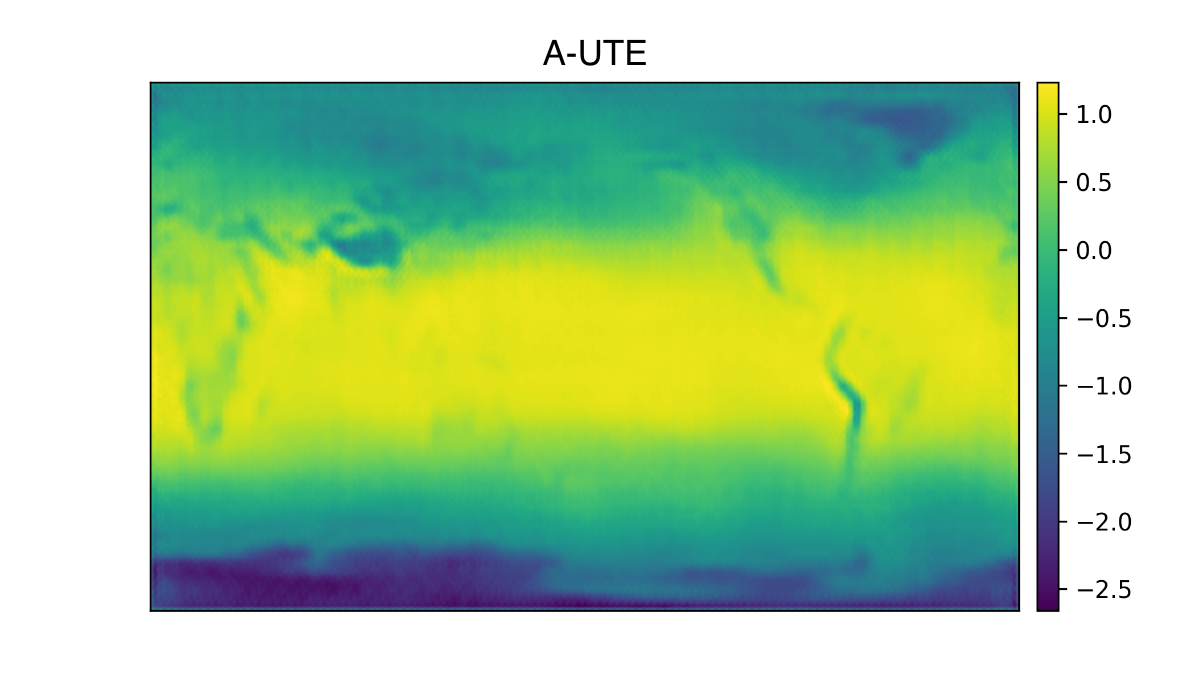}
    \end{subfigure}
    \caption{Time–mean maps for air temperature fields across vertical levels (2,3,4,5). Both the quantitative results in table \ref{tab:fv3gfs} and visual results indicate stronger fidelity at higher levels (L4–L5) than at lower tropospheric levels (L2–L3), with reduced discretizations.} 
    \label{fig:fv3maps}
\end{figure*}

Table~\ref{tab:fv3gfs} reports 10-year roll-outs on FV3GFS at six vertical
levels. A-UTE attains the lowest RMSE on all six levels and the lowest CRPS on five of six. At the near-surface level $T_1$, NADE's empirical CRPS (0.342) is marginally below A-UTE's (0.347, a gap of 0.005\,K), while A-UTE's RMSE on $T_1$ remains lower. Figure~\ref{fig:fv3maps} shows time-mean reference and emulated temperature fields at intermediate levels, with agreement visibly tighter at the upper levels than at the lower tropospheric levels, matching the quantitative ordering.

\begin{table*}[!h]
\centering
\caption{10-year auto-regressive roll-outs on FV3GFS at six vertical levels. Latitude-weighted RMSE (K) and CRPS (K); lower is better. CRPS for the deterministic baselines is computed from a 5-seed empirical ensemble. Best entry per metric in bold.}
\label{tab:fv3gfs}
\small
\setlength{\tabcolsep}{4pt}
\begin{tabular}{l cc cc cc cc cc}
\toprule
& \multicolumn{2}{c}{A-UTE} & \multicolumn{2}{c}{U-Net} & \multicolumn{2}{c}{ConvLSTM} & \multicolumn{2}{c}{SFNO} & \multicolumn{2}{c}{NADE} \\
\cmidrule(lr){2-3}\cmidrule(lr){4-5}\cmidrule(lr){6-7}\cmidrule(lr){8-9}\cmidrule(lr){10-11}
Level & RMSE & CRPS & RMSE & CRPS & RMSE & CRPS & RMSE & CRPS & RMSE & CRPS \\
\midrule
$T_1$ & \textbf{0.437} & 0.347 & 0.924 & 0.876 & 0.653 & 0.598 & 0.466 & 0.354 & 0.452 & \textbf{0.342} \\
$T_3$ & \textbf{0.408} & \textbf{0.274} & 0.865 & 0.772 & 0.656 & 0.582 & 0.482 & 0.392 & 0.492 & 0.398 \\
$T_4$ & \textbf{0.355} & \textbf{0.222} & 0.796 & 0.622 & 0.592 & 0.462 & 0.491 & 0.365 & 0.382 & 0.323 \\
$T_5$ & \textbf{0.349} & \textbf{0.221} & 0.899 & 0.731 & 0.536 & 0.401 & 0.379 & 0.399 & 0.421 & 0.392 \\
$T_6$ & \textbf{0.379} & \textbf{0.240} & 0.738 & 0.598 & 0.582 & 0.421 & 0.459 & 0.387 & 0.451 & 0.372 \\
$T_7$ & \textbf{0.362} & \textbf{0.237} & 0.729 & 0.601 & 0.562 & 0.398 & 0.381 & 0.299 & 0.452 & 0.394 \\
\bottomrule
\end{tabular}
\end{table*}

\subsection{Zero-Shot Cross-Model Emulation}
\label{sec:zeroshot}
Zero-shot evaluation assesses whether the dynamics learned on one model transfer, without retraining, to a structurally different model. Climate models differ in their numerical schemes, parameterisations, and mean-state biases, which induce systematic shifts in spatial spectra and temporal variability, so robust transfer indicates that the learned dynamics capture transport- and forcing-consistent spatiotemporal structure. We pre-train on each of three source models (AWI-CM-1-1-MR, EC-Earth3, TaiESM1) and evaluate, without fine-tuning, on distinct target models at the 1-year roll-out horizon (Table~\ref{tab:zeroshot}). The 1-year horizon is used here because at 10 years the cumulative effect of cross-model bias drift dominates and obscures the immediate transfer behaviour. The 10-year evaluation in Section~\ref{sec:forced} is the appropriate test of in-distribution long-horizon stability. A-UTE attains the lowest RMSE and CRPS across the source$\rightarrow$target pairs, and cross-model RMSE stays within roughly 10\% of the in-domain values, indicating that its representation of monthly temperature evolution is not strongly conditioned on model-specific signatures.

\begin{table*}[!h]
\centering
\caption{Zero-shot cross-model emulation at the 1-year roll-out horizon. Each block lists the source model in its first row (in-domain) and target models evaluated without fine-tuning. RMSE (K) and CRPS (K); lower is better. CRPS for the deterministic baselines reflects the 5-seed empirical construction. Best entry per metric in bold.}
\label{tab:zeroshot}
\small
\setlength{\tabcolsep}{4pt}
\begin{tabular}{l cc cc cc cc cc}
\toprule
& \multicolumn{2}{c}{A-UTE} & \multicolumn{2}{c}{U-Net} & \multicolumn{2}{c}{ConvLSTM} & \multicolumn{2}{c}{SFNO} & \multicolumn{2}{c}{NADE} \\
\cmidrule(lr){2-3}\cmidrule(lr){4-5}\cmidrule(lr){6-7}\cmidrule(lr){8-9}\cmidrule(lr){10-11}
Target & RMSE & CRPS & RMSE & CRPS & RMSE & CRPS & RMSE & CRPS & RMSE & CRPS \\
\midrule
\multicolumn{11}{l}{\textit{Source: AWI-CM-1-1-MR}} \\
AWI-CM-1-1-MR     & \textbf{0.190} & \textbf{0.134} & 0.534 & 0.423 & 0.589 & 0.479 & 0.322 & 0.210 & 0.298 & 0.182 \\
EC-Earth3         & \textbf{0.214} & \textbf{0.149} & 0.759 & 0.598 & 0.590 & 0.481 & 0.342 & 0.230 & 0.310 & 0.230 \\
GFDL-ESM4         & \textbf{0.198} & \textbf{0.136} & 0.603 & 0.502 & 0.599 & 0.489 & 0.325 & 0.213 & 0.242 & 0.139 \\
FGOALS-f3-L       & \textbf{0.225} & \textbf{0.155} & 0.659 & 0.553 & 0.587 & 0.477 & 0.351 & 0.240 & 0.336 & 0.215 \\
\midrule
\multicolumn{11}{l}{\textit{Source: EC-Earth3}} \\
EC-Earth3         & \textbf{0.193} & \textbf{0.136} & 0.606 & 0.498 & 0.594 & 0.479 & 0.406 & 0.304 & 0.276 & 0.195 \\
TaiESM1           & \textbf{0.216} & \textbf{0.149} & 0.569 & 0.421 & 0.599 & 0.486 & 0.464 & 0.352 & 0.364 & 0.252 \\
CNRM-CM6-1-HR     & \textbf{0.197} & \textbf{0.138} & 0.591 & 0.436 & 0.610 & 0.511 & 0.476 & 0.364 & 0.299 & 0.175 \\
INM-CM4-8         & \textbf{0.214} & \textbf{0.146} & 0.581 & 0.425 & 0.598 & 0.499 & 0.495 & 0.383 & 0.325 & 0.210 \\
\midrule
\multicolumn{11}{l}{\textit{Source: TaiESM1}} \\
TaiESM1           & \textbf{0.193} & \textbf{0.136} & 0.633 & 0.511 & 0.576 & 0.466 & 0.432 & 0.387 & 0.311 & 0.201 \\
INM-CM5-0         & \textbf{0.194} & \textbf{0.138} & 0.661 & 0.530 & 0.617 & 0.509 & 0.479 & 0.391 & 0.312 & 0.211 \\
MPI-ESM1-2-HR     & \textbf{0.205} & \textbf{0.144} & 0.646 & 0.512 & 0.580 & 0.476 & 0.461 & 0.350 & 0.332 & 0.220 \\
NorESM2-MM        & \textbf{0.202} & \textbf{0.140} & 0.657 & 0.521 & 0.579 & 0.456 & 0.471 & 0.359 & 0.377 & 0.253 \\
\bottomrule
\end{tabular}
\end{table*}

\subsection{Ablation: Role of the Advection Prior}
\label{sec:ablation}
\begin{figure}[!h]
    \centering
    \includegraphics[width=\linewidth]{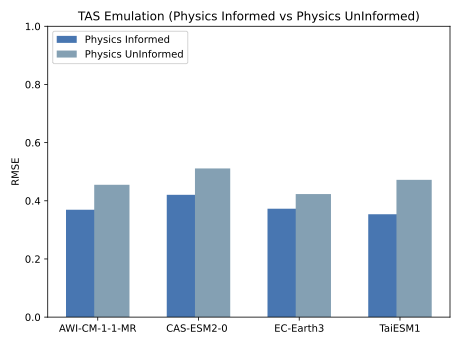}
\caption{Ablation of the advection-forcing core across four CMIP6 models. Latitude-weighted RMSE for 10-year roll-outs comparing the full A-UTE (physics-informed) against the variant with the transport core removed and only the latent neural SDE retained (physics-uninformed). Removing the core raises RMSE on all four models, with the largest degradation on CAS-ESM2-0 and TaiESM1.}
\label{fig:ablation}
\end{figure}

To isolate the contribution of the advection-forcing core, we remove it entirely and retain only the latent neural SDE, leaving a purely data-driven stochastic emulator. Figure~\ref{fig:ablation} reports 10-year RMSE across four representative ClimateSet models for the full model and the ablated variant. RMSE increases on all four when the advection core is removed, with the largest degradation on CAS-ESM2-0 and TaiESM1. The transport-shaped prior therefore contributes substantively to skill, and the unconstrained latent SDE does not recover the lost accuracy within the same training budget, indicating that the deterministic core steers the emulator toward a function class the SDE alone does not reach.

\subsection{Computational cost}
A-UTE is trained once per configuration in roughly 6 days on a single NVIDIA H100 GPU. Once trained, it emulates a full 10-year (120-step) monthly roll-out, including the $5$ Brownian samples that form the predictive distribution, in about a minute on the same GPU on the order of $10^4$ simulated years per day. For comparison, CMIP6-class Earth system models at a comparable resolution require on the order of $10^3$-$10^4$ core-hours per simulated year on CPU clusters. Among the models emulated here, MPI-ESM1 and EC-Earth3 have been reported at roughly 3,400 and 12,000 core-hours per simulated year, respectively \cite{balaji2017cpmip}. 

\section{Discussion}
\label{sec:discussion}

The results in Section~\ref{sec:results} support three claims. A-UTE attains lower RMSE and CRPS than four learning-based baselines across 19 ClimateSet models and six FV3GFS vertical levels over 10-year roll-outs (Sections~\ref{sec:forced}). The learned dynamics transfer across structurally different models in 1-year zero-shot evaluation (Section~\ref{sec:zeroshot}), and the advection-forcing core contributes substantively to skill (Section~\ref{sec:ablation}). Relative to the emulator families surveyed in Section \ref{sec:related}, A-UTE occupies the monthly-resolution, multi-model, scenario-conditioned regime, pairing a transport-shaped spatial prior with likelihood-based training that exposes a calibrated per-grid-cell predictive distribution. We discuss what these properties imply for deploying the emulator as a practical tool in spatial climate-impact workflows.

\subsection{From Application to Deployment}
\label{sec:deployment}

The value of a temperature emulator in practice is set by three properties: how cheaply it produces gridded fields, how well it generalises across the models and scenarios an analyst needs, and whether it reports trustworthy uncertainty. A-UTE is designed around these properties, which positions it for deployment as a fast surrogate inside the spatial workflows that currently depend on full Earth system model (ESM) output.

\paragraph{Ensemble-scale scenario exploration.}
Because A-UTE conditions directly on time-varying forcing maps and emulates the gridded monthly field by forward integration, it can generate large ensembles of spatially resolved temperature trajectories under arbitrary prescribed pathways at a small fraction of the cost of a coupled ESM integration \cite{balaji2017cpmip}. This is the capability that makes routine, many-member, many-scenario screening feasible: an analyst can sweep forcing pathways and initial conditions to characterise both the forced response and its spread, then reserve expensive ESM runs for the configurations that matter most \cite{tebaldi2025emulators, van2026rewiring}.

\paragraph{One surrogate across the model archive.}
The zero-shot transfer behaviour in Section~\ref{sec:zeroshot} indicates that the learned operator captures transport- and forcing-consistent structure that holds across structurally different simulators rather than memorising one model's signature. For deployment this means a single trained emulator can serve as a fast stand-in across much of the CMIP archive \cite{kaltenborn2023climateset}, extending ensemble size and model diversity together without retraining a bespoke emulator per source model.

\paragraph{Calibrated uncertainty for decision support.}
Many downstream spatial analyses consume not only the most likely temperature field but the confidence attached to it. A-UTE reports a heteroscedastic predictive variance at every grid cell, trained under a likelihood objective and evaluated probabilistically through CRPS. In a deployed setting this supports risk-aware use of the emulated fields, where local predictive spread feeds directly into heat-exposure mapping, degree-day calculations, and other impact models that the established statistical emulators already drive \cite{beusch2020emulating, nath2022mesmer}.

\paragraph{Pathway and prerequisites.}
The present evaluation is offline, against held-out model trajectories. The emulator is not yet embedded in an operational impact pipeline. A realistic deployment route would expose A-UTE as a forcing-conditioned emulation service that returns gridded monthly temperature with uncertainty for a requested scenario, feeding existing impact tooling. Reaching that point calls for a measured wall-clock characterisation of the inference cost against the ESM baseline, a reliability analysis of the predictive intervals under operational conditions, and training at the higher spatial resolutions that regional impact applications require. These are concrete and tractable steps, and the properties demonstrated here are the ones a deployed surrogate would rely on.

\section{Conclusion, Limitations, and Future Work}
\label{sec:conclusion}

\paragraph{Conclusion.}
We presented A-UTE, an advection-informed, uncertainty aware emulator for spatially resolved monthly-mean surface temperature. A-UTE couples a two-dimensional forced-advection ODE, acting as a transport-shaped spatial prior driven by a data-derived pseudo-velocity field, with a latent neural SDE residual trained under a latitude-weighted Gaussian negative log-likelihood. Across 19 CMIP6 models from ClimateSet and the FV3GFS model, A-UTE achieves lower RMSE and CRPS than U-Net, ConvLSTM, SFNO, and a graph-based advection-diffusion baseline over 10-year roll-outs, under a cross-scenario split that trains on SSP1-2.6 and tests on SSP2-4.5. The learned dynamics transfer to unseen models at the 1-year zero-shot horizon, and an ablation of the advection core confirms that the transport-shaped prior contributes substantively to skill. More broadly, the results indicate that embedding a modest, physically motivated spatial prior into a likelihood-trained dynamical emulator is a productive design choice for monthly, scenario-conditioned, multi-model temperature emulation.

\paragraph{Limitations.}
The advection prior represents one term of the monthly heat budget. Equation \eqref{eq:pde} models horizontal transport, while the monthly tendency of surface temperature also reflects radiative forcing, surface energy fluxes, latent-heat release, and turbulent mixing, which are not included as explicit source terms. As a result, the pseudo-velocity in \eqref{eq:speed} subsumes these processes. It acts as a data-derived inductive bias rather than a physical wind field, encoding upper-level planetary-wave dynamics alongside near-surface flow. The ablation in Section~\ref{sec:ablation} shows that this bias improves skill, and a comparison of the inferred pseudo-velocity against the true circulation would clarify what it captures. The forcing set for the ClimateSet experiments omits incoming solar radiation, so the seasonal cycle is learned implicitly from the temperature field. The FV3GFS configuration includes the downward shortwave flux explicitly. The evaluation reports aggregated rather than per-month metrics and extends to a 10-year horizon, longer integrations remain to be characterised. Finally, the CRPS comparison constructs an empirical predictive distribution for the deterministic baselines, which credits them with a spread they were not trained to produce, and the emulator is trained on the coarse ClimateSet and FV3GFS grids, which do not resolve regional extremes or sub-grid orographic effects.

\paragraph{Future work.}
We identify three primary directions for extending this framework:
\textbf{Architectural extensions:} Expanding the core ODE to explicitly include energy-budget contributions such as radiative forcing and surface-flux divergence would clarify whether performance gains stem specifically from transport mechanics or simply from having a physically motivated prior. Additionally, the architecture naturally extends to the joint emulation of precipitation and near-surface humidity \cite{schongart2024introducing}.
\textbf{Rigorous benchmarking:} To clearly position A-UTE against established methodologies, a harmonized head-to-head comparison with the MESMER family \cite{nath2022mesmer} is essential, with a particular focus on the variability spectra required for downstream applications.
\textbf{Operational impact capabilities:} Coupling our atmosphere-only emulator with a simple ocean representation would allow for dynamic $CO_2$-sensitivity experiments rather than relying on fixed-trajectory SSP runs \cite{clark2025ace2}. Paired with higher-resolution training, evaluation on regional extremes, and multi-decadal roll-outs \cite{guan2025lucie, cresswell2025deep}, these improvements would fully bridge the gap to operational impact assessment.

\begin{acks}
This work was supported by SmartSAT Cooperative Research Center (project number P3-31s). We would also like to acknowledge National Computational Infrastructure (NCI), a high performance computing center and ResetData (\url{https://resetdata.com.au/}) for providing us with GPU resources and data collection which enabled us to perform this research.
\end{acks}

\section*{Ethics and Privacy Statement}
This work trains a temperature emulator on public, documented climate-model output \cite{kaltenborn2023climateset, eyring2016overview, zhou2019toward}, used under the relevant data licenses. It involves no human-subject data or personally identifiable information, so institutional-review privacy concerns do not arise. The importance of A-UTE lies in generating computationally efficient, uncertainty-aware temperature projections which can widen access to ensemble-scale scenario analysis and reduce the energy cost of climate computation. The principal risk associated with this work is the potential over-reliance on, or misuse of, the emulated fields. Since, A-UTE is trained on coarse-resolution data and approximates complex physical processes, its outputs may exhibit regionally heterogeneous errors. Therefore, these projections should complement rather than replace validated physics-based model outputs. To mitigate this risk and encourage responsible downstream use, our framework explicitly outputs calibrated per-grid-cell uncertainty. We conclude that the broader impacts and accessibility benefits of this approach substantially outweigh the identified risks, provided the tool is deployed with a clear understanding of its documented limitations.

\bibliographystyle{ACM-Reference-Format}
\bibliography{ref}

\end{document}